# Deep Learning Methods for Software Requirement Classification: A Performance Study on the PURE dataset

Fatemeh Khayashi, Behnaz Jamasb, Reza Akbari, Pirooz Shamsinejadbabaki

Department of Computer Engineering and Information Technology, Shiraz University of Technology, Shiraz, Iran

---

**Abstract**: Requirement engineering (RE) is the first and the most important step in software production and development. The RE is aimed to specify software requirements. One of the tasks in RE is the categorization of software requirements as functional and non-functional requirements. The functional requirements (FR) show the responsibilities of the system while non-functional requirements represent the quality factors of software. Discrimination between FR and NFR is a challenging task. Nowadays Deep Learning (DL) has entered all fields of engineering and has increased accuracy and reduced time in their implementation process. In this paper, we use deep learning for the classification of software requirements. Five prominent DL algorithms are trained for classifying requirements. Also, two voting classification algorithms are utilized for creating ensemble classifiers based on five DL methods. The PURE, a repository of Software Requirement Specification (SRS) documents, is selected for our experiments. We created a dataset from PURE which contains 4661 requirements where 2617 requirements are functional and the remaining are non-functional. Our methods are applied to the dataset and their performance analysis is reported. The results show that the performance of deep learning models is satisfactory and the voting mechanisms provide better results.

---

## 1. Introduction

Developing high-quality software is an expensive and time-consuming task. Automating software engineering tasks in different phases such as requirements engineering, analysis, design, and implementation helps us to develop software in a shorter time with better quality [1]-[5]. Determining software requirements is the first and most important part of the software engineering process. Software requirements are services that software must provide and include the restrictions that are applied to these services. In general, the services that must be provided by the software are called FR, and the quality of presenting these services lay in NFRs such as usability, reliability, security, performance, etc.

Software users express their needs in the natural language form and usually verbosely. From those ambiguous and unstructured data, an analyst must extract the requirements of the software in a way that the designer can understand them. Due to the ambiguity of natural language, it is difficult to extract each requirement accurately by humans. It is obvious that if the software's FRs and NFRs are not extracted correctly, the software development project can not meet the goals of the end users. Therefore, the correct and complete extraction of FRs and NFRs in a short time and at the beginning of the software production cycle is very important. To classify requirements in a more accurate and faster way, the development team can use Artificial Intelligence (AI) methods. Today, Machine Learning (ML) and DL methods have transformed computers into intelligent and hard-working assistants of humans in different domains like medicine, industry, and even software engineering. Hence, It seems that these methods can be applied successfully to classify software requirements.

Since the discrimination of FRs and NFRs manually is a difficult and time-consuming task, this work is aimed to present a solution based on DL methods for the classification of software requirements. For this purpose, the PURE dataset is used which is a collection of SRSs of different types. The PURE's SRSs have been provided in different formats and styles. Hence, we need to generate a dataset for our algorithms based on PURE. The generated dataset has 4661 requirements. The dataset is preprocessed and then analyzed by five DL algorithms along with two voting methods. The algorithms showed competitive performance. However, the best results are obtained by voting methods.

The main contributions of the paper are:

- Creating a dataset from the PURE repository for requirement classification.
- Applying some of the most prominent DL methods for requirement classification.
- Creating ensemble models from DL methods.

The remaining of this paper is organized as follows: the next section presents the related work. Section 3 contains the details of the dataset preparation and the proposed methods. The performance of the DL methods is given in Section 4. Finally, Section 5 concludes this work.

## 2. Related Work

In recent years, researchers in the software engineering domain have extensively used AI methods to cope with the requirement engineering challenges. Generally, the requirements can be divided into functional and non-functional. Categorizing the software requirements as functional and non-functional helps the analyst and designer to better understand the software and provide more precise SRS.

Almanza et al. used Convolutional Neural Network (CNN) to classify NFRs [6]. They applied their model to the PROMISE dataset and classified NFRs into 11 categories of availability, security, maintainability, performance, efficiency, scalability, usability, fault tolerance, portability, permission, and appearance features. They expressed the results with three criteria: F-score, recall, and precision.

Rahman et al. classified the NFRs using Long Short-Term Memory (LSTM), Gated Recurrent Unit (GRU), and CNN [7]. The PROMISE dataset has been used, which contains 625 necessary sentences, where 255 are functional and 370 are non-functional requirements. The results showed that the LSTM algorithm performed better than the other two algorithms. Winkler and Vogelsang categorized software requirements into two categories: requirements and information using CNN [8]. They used the word2vec technique for data conversion and recall, precision, and f-score criteria to evaluate the performance. They used DOORS dataset which contains 89 documents.

Baker et al. [4] used Artificial Neural Networks (ANN) and CNN algorithms [9]. First, they divided the software requirements into FRs and NFRs. After that, the NFRs have been classified into four categories: usability, security, operational, and efficiency. The dataset used in this work is a collection of user opinions about software.

Rahimi et al. [10] used the PROMISE dataset. Their work includes two phases: In the first phase, they divided the set of software requirements into FRs and NFRs. In the second phase, they classified functional requirements into six categories: solution, activation, attribute limitation, action limitation, policy, and definition; and NFRs into 11 categories of availability, security, maintainability, performance, efficiency, scalability, usability, fault tolerance, portability, permission, and appearance characteristics. They used GRU, Bidirectional LSTM (BiLSTM), LSTM, and CNN, and their ensembles called ensemble average, accuracy as weight, and accuracy in each class as weight. The best results were obtained by ensembles.

A new model based on Bidirectional Encoder Representations from Transformers (BERT) called DistilBERT has been presented by Kici et al. [11]. They used the DOORS dataset and placed the SRS documents in this dataset into three categories: type, priority, and severity. They determined that there were 21 classes for the type attribute. Also, for the priority attribute, there are 4 classes, and for the severity attribute, there are 5 classes. They classified these three features separately using DistilBERT, LSTM, BiLSTM, BiLSTM+GloVe, and BiLSTM+Word2vec algorithms. For all three classifications, the proposed DistilBERT algorithm had the best performance. They also conducted their experiment on the PROMISE dataset, where the DistilBERT algorithm performed best.

Ivanov et al. used the PURE dataset [12]. They manually extracted the information contained in 79 documents of the PURE dataset and divided them into two categories of requirements and non-requirements. Then they did their experiment with three models: Support Vector Machines (SVM)+ (ELMo) Embedding from Language Models, BERT, and FastText. The results of their experiment show that the BERT model had the best performance. They also tested their model on the RFI dataset, which contains 380 sentences, and the results show that the BERT algorithm performed better on this dataset as well.

Malik et al. identified nominal entities in the SRS [13]. They used the DOORS dataset and labeled 12 groups of nominal entities called non-institutional tokens, core, user, graphical user interface, hardware, language, application programming interface, standard, platform, adjective, and verb. Then they used BiLSTM-CRF, MEM, ML-(CRF) Conditional Random Field (CRF), and BiLSTM-CRF (GloVe) to classify nominal entities. The results showed that the ML-CRF model had the best performance.

Rahimi et al. [14] classified the FRs into 6 classes: solution, empowerment, action limitation, feature limitation, definition, and policy. The dataset used in this work includes 600 FRs, where each class contains 100 requirements. They used 5 ML algorithms: Naïve Bayes (NB), SVM, Decision Tree (DT), Linear Regression (LR), and Support Vector Machine (SVC). They have also used two methods of Term Frequency-Inverse Document Frequency (TF-IDF) feature extraction and vector counting.

Tiun et al. applied the combination of SVM, NB, and LR with Doc2Vec, and also they used the combination of Word2vec and CNN, and the FastText algorithm on the PROMISE dataset [15]. The results showed that the FastText method performed best. Also, they have used the combination of SVM, NB, and LR algorithms with Bag of Words (BoW), as well as the combination of SVM, NB, and LR algorithms with TF-IDF, and the combination of CNN and Word2Vec algorithms. The results showed that the combination of LR and TF-IDF had the best performance, which shows that DL algorithms do not necessarily give better results than traditional algorithms in text problems with short length.

Haque et al. [16] used Multinomial Naïve Bayes (MNB), Gaussian Naïve Bayes (GNB), Binarized Naïve Bayes (BNB), K-Nearest Neighborhood (KNN), SVM, SGD SVM, and DT algorithms to classify NFRs [11]. They used BoW and TF-IDF at character, word, and n-gram levels to extract features and convert text to vector. The results showed that the SGD SVM algorithm had the best performance.

Shakeri et al. divided the software requirements into FR and NFR [17]. Then the NFRs were divided into 10 categories. They used a special preprocessing method. Then, using the C4.5 algorithm, they separated the FR and NFR of the software once for preprocessed data and once for non-preprocessed data. The results showed that the proposed method had better performance on preprocessed data. They have used Biterm Topic Model (BTM), Hierarchical, Hybrid, K-means, and Binarized Naive Bayes (BNB) algorithms. The results showed that the BNB algorithm performed best both in preprocessed and non-preprocessed data. Also, the preprocessing had a significant effect. The BTM algorithm is better than Latent Dirichlet Allocation (LDA), but here LDA algorithm has better results than the BTM algorithm. They used the PROMISE dataset.

Lu et al. extracted the requirements from the users' opinions about some apps in Google Play and AppStore [18]. They have divided users' opinions into FR, NFR, and other items. First, they manually labeled data and determined which category each opinion belongs to. Then they preprocessed the data. After that, CHI2, TF-IDT, BoW, and AUR-BoW algorithms were used to convert text to vector, and then J48, NB, and Bagging algorithms were used separately for classification. The results showed that AUR-BoW along with the Bagging technique and C4.5 obtained the best performance.

In [19] Jindal et al. classified software security requirements into 5 categories using the J48 algorithm. In this work, after preprocessing the data, they used the TF-IDF algorithm to extract the features, then according to the large number of extracted features, they used the Info-Gain criterion to reduce the number of features. They used the PROMISE dataset of 58 security-based descriptions extracted from a total of 15 projects.

Canedo et al. [20] used the PROMISE dataset and initially classified requirements into FR and NFR. Then they classified NFRs into 11 subcategories, and FRs into 12 subcategories. They intended to conclude which combination of feature extraction method with ML algorithm works better. They used three methods: TF-IDF, CHI2, and BoW for feature extraction. After that, four algorithms LR, KNN, MNB, and SVM were used. The results showed that the combination of the TF-IDF method with the LR algorithm had the best performance.

In [21], Deocadez et al. used the semi-supervised algorithms RAndom Subspace Method for Co-training (RASCO), self-training, and Relevant Random Subspace Co-training (Rel-RASCO) algorithms to classify software requirements into FR and NFR. In this work, 300 reviews of mobile software in AppStore were used. Half of them are FRs and the other half are NFRs. They used the above algorithms, as well as four NB, C4, 5, SVM, and KNN algorithms as the base algorithms for the three semi-supervised algorithms.

Talele and Phalnikar [22] used PURE datasets and a dataset of user comments about Amazon products, and after preprocessing, they used TF-IDF, BoW, and CHI2 methods to extract features; and SVM, DT, KNN, LR, NB, and MNB algorithms to classify requirements into FR and NFR. The results showed that SVM and decision tree algorithms performed best. In this work, the performance of software requirements prioritization algorithms has also been investigated.

Toth and Vidacs [23] classified NFRs in the PROMISE dataset. After pre-processing and feature extraction using the TF-IDF method, they used KNN, ETs, GNB, ET, DT, label spread, label spread, Logistic, SVM, MNB, and MLP algorithms to classify non-functional requirements. Their results showed that MNB, SVM, and linear logistic regression (LLR) algorithms have performed best. They also considered the execution time, and by considering it, they found that the MNB algorithm had the best performance.

In [24], Toth and Vidacs identified and classified the NFRs. They conducted two experiments. In the first experiment, they used the PROMISE small dataset, and KNN, GNB, ET, DT, BNB, label propagation, expansion Labels, SVM, MNB, MLP, and LR algorithms were used for classification. In the second experiment, they used a large dataset collected from Stack Overflow; They also used SVM, KNN, LR, ET, DT, MNB, GNB, BNB, and fully connected network algorithms for classification. The results showed that linear classification algorithms performed best in both sets. SVM, MNB, and LR algorithms had the best performance, although the DT algorithm also performed relatively well. The fully connected network also had a good result in the second test.

Anas and Williams [25] used an unsupervised approach to identify, classify and track NFRs. They used the semantic similarity of words of FRs to find and cluster NFRs; Because they believe that NFRs are implicitly expressed along with FRs. They found that the NGDWiki and LSA methods perform better than the PMI method. In general, NGDWiki does not have many vocabulary problems because it is based on Wikipedia. They found that hierarchical clustering methods perform better than partitioning algorithms.

Table 1 shows a taxonomy of the methods proposed for software requirements classification. It shows the objective of the work, class of algorithm used, feature extraction method, evaluation metric, dataset, and the best algorithm(s). Most of the works used the PROMISE dataset. A few works considered the PURE dataset.

Table 1: A taxonomy of the methods used for requirement classification

| Ref. | Objective | Algorithm | Data Set | Best algorithm |
|------|-----------|-----------|----------|----------------|
| [6] | FR classification | DL | PROMISE | CNN |
| [7] | NFR classification | DL | PROMISE | LSTM |
| [8] | Categorizing requirements as requirements and information | DL | DOORS | CNN |
| [9] | Division of the software requirements as FR and NFR, and classification of NFR | DL, ML | PROMISE | CNN |
| [10] | Classification of requirements as FR and NFR. Sub-classification of FR and NFR | DL, ML | PROMISE | Ensemble models |
| [11] | Software requirements classification | DL | DOORS, PROMISE | DistilBERT |
| [12] | Classification of SRS as requirements and non-requirements | ML | PURE, RFI | BERT |
| [13] | Classification of nominal entities in the SRS | DL | DOORS | ML-CRF |
| [14] | FR Classification | ML | Dataset of 600 FR | Ensemble models |
| [15] | Classifying requirements into FR and NFR. | ML | PROMISE | LR |
| [16] | Classification of NFR | ML | PROMISE | SGD SVM |
| [17] | Classifying requirements into FR and NFR. Classification of NFR | ML | PROMISE | BNB |
| [18] | Classification of users' opinions as FR, NFR, and others. | ML | Whatsapp, iBook | ARU-BoW, C4.5 |
| [19] | Classification of security requirements | ML | PROMISE | - |
| [20] | Classification of requirements as FR and NFR. Sub-classification of FR and NFR | ML | PROMISE | LR |
| [21] | Classification of requirements as FR and NFR | ML | User reviews | |
| [22] | Classification and prioritization of software requirements | ML | PURE, user comments | SVM and DT |
| [23] | NFR classification | ML | PROMISE | MNB, SVM LLR, MNB |
| [24] | NFR classification | ML | PROMISE, Stack Overflow | SVM, MNB, and LR. |
| [25] | NFR classification | ML | 3 software req. | NGDWiki and LSA |

## 3. Proposed Method

The structure of the proposed method is given in Figure 1. The method has two phases: preparing the dataset, and designing prediction methods. The second phase has three steps: preprocessing, feature extraction, and classification using DL methods.

### A) Dataset Preparation

The PURE is a repository of 79 SRS of different types. 63 SRSs are pdf, 14 of them are word documents, and 3 of them are Html. The type of documents is given in Table 2. The documents in the PURE are named System requirement specification, Software requirement specification, Functional requirement specification, or Functional & non-functional requirement specification. For the sake of simplicity, we refer to all of them as SRS in this work. The shortest SRS has 8 pages while the largest one has 288 pages.

Table 2. Types of documents in the PURE repository.

| Type | # |
|---|---|
| System requirement specification | 10 |
| Software requirement specification | 48 |
| Functional requirement specification | 19 |
| Functional & non-functional requirement specification | 1 |
| Undefined | 1 |
| Total | 79 |

The learning algorithms adapted in this work cannot use the repository for prediction directly. Hence, it is necessary to process the SRSs in different formats and create a suitable dataset. For this purpose, we have done different tasks. As shown in Table 1, previous works on requirement classification had different goals. Some works tried to classify the requirements as FR and NFR, while some other works tried to classify NFR into quality factors. A few works aimed to discriminate between information and requirements. Finally, a few works tried to classify FR.

Here, we focus on the classification of software requirements as FR and NFR. Hence, the class label is determined as "Functional Requirement" and "Non-Functional Requirement". To create the dataset, we need to do two main tasks: investigating the SRSs to find the requirements, and determining the class label. These tasks have been done manually. The SRSs have been analyzed manually and software requirements are extracted from them. In some SRSs, software requirements are clearly described and it is straightforward to extract them. However, in some cases, you need to pay more attention to extract the requirements. The requirements have different lengths. Some of them are short sentences while the other ones are represented as paragraphs containing two or more lines.

In some SRSs, the class labels of the requirements are given (a requirement can be categorized as functional or non-functional). However, in some cases, the type of requirements are not given. In such cases, we have analyzed the requirements, and have determined the labels. In some cases, the NFRs are given as quality factors such as security, reliability, performance, etc. All the quality requirements are considered as NFR.

By processing the SRSs we have a dataset containing 4661 records where 2617 requirements are functional and 2044 requirements are non-functional. Each record presents the text of a requirement and its class label. A sample of extracted FR and NFR is presented in Table 3.

Table 3. A typical FR and NFR from the PURE repository.

| Requirement | Class |
|---|---|
| "System provides a management console displaying workstations running client software; workstation name and IP address; and utilities for managing client sessions." | FR |
| "The Application should be available always at working hours. Any maintenance or backup operation should be conducted out of working time" | NFR |

### B) Preprocessing

After generating the dataset from the PURE repository, we are ready to apply the DL methods to classify the requirements. Figure 1 shows the structure of the proposed method. The proposed method has three steps. The first step which is known as preprocessing starts by collecting and specifying the dataset. At this stage, the input data is cleaned and purified. Since the input data is text, the set of steps we have taken to preprocess the data are as follows. To do this, we have used the NLTK library in Python programming language:

- Removing numbers and punctuation marks.
- Converting uppercase letters to lowercase.
- Eliminating stop words: in English, words like the, a, etc are called stop words. These words should be removed because they have no value or meaning in the sentence.
- Rooting: In this step, the root of each word replaces that word. Because words with the same root are synonymous and only differ from each other in terms of appearance.

### C)  Feature Extraction

In the second step, the features of each requirement should be extracted. There are different ways for feature extraction. In previous works, methods such as TF-IDF, Word2Vect, BoW, word embedding, etc. have been used for feature extraction. Here, we used the word embedding method which displays each word with a vector of numbers in such a way that words that are conceptually related to each other have embedding vectors close to each other.

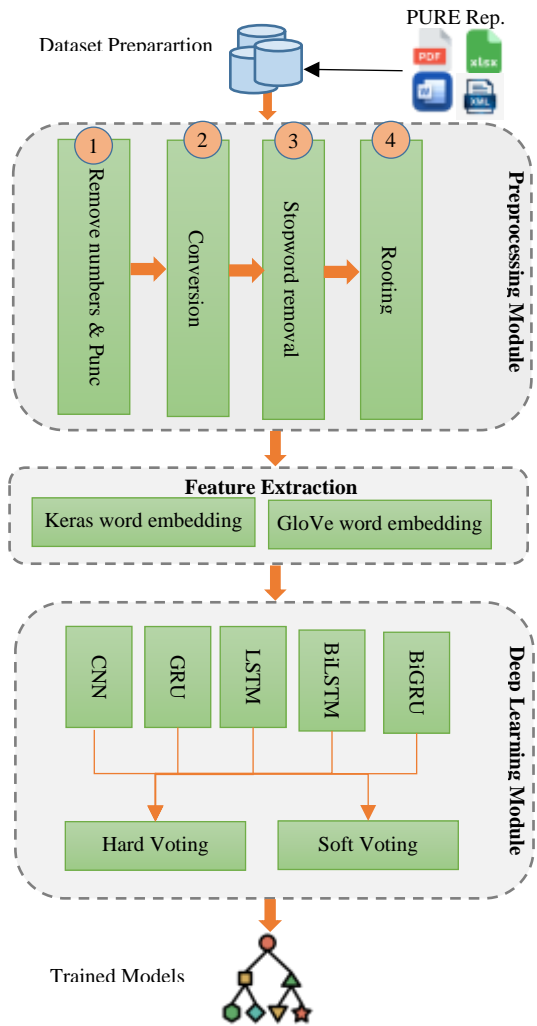

Figure 1. The overall structure of the proposed method

Here, we have used two word embedding techniques, one is available in the Keras library and is trained using the dataset of the problem. The other is a word embedding from Pre-trained GloVe trained on the Wikipedia dataset. GloVe is an unsupervised learning algorithm to obtain vector reprentations of the words. The numerical vector otained for each word contains 300 dimensions, that is, it defines each word with 300 features. The vector represenations of athe words is obtained by mapping words into a space where the semantic similarity between words is considered as the distance.

### D)  Deep Learning Module

In this step, the dataset has been divided into training and testing sets. The extracted features of the training set are fed to the learning methods to learn the problem. After that, the learned methods are used to classify the requirements. The output of the proposed method is a set of trained DL and voting models which can be used to classify software requirements into two categories of FRs and NFRs.

In recent years, different DL methods have been proposed by researchers and successfully applied to different problems. In this work, we use five DL methods: LSTM, BiLSTM, GRU, BiGRU, and CNN. In addition, two voting methods known as hard voting and soft voting are used. The model is trained using each DL algorithm separately. Then the voting methods have been leveraged for predicting the final class of each requirement in the test dataset.

## 4.  Performance study

It seems that the proposed method has the abality to perform well in classifying software requiremets. This section presents the results obtained by the DL and voting methods.

### A)  Experimental Settings

We used Python 3.8 and PyCharm IDE to develop the proposed method. Six libraries were used. The Pandas were used to import data from the dataset. The NLTK library was used to preprocess the data. The Scikit-Learn was used for data partitioning and evaluation metrics. Finally, Tensorflow and Keras libraries were used for tokenization, padding, DL methods, etc.

To evaluate the performance of the proposed model, we need to determine the parameters of the DL methods along with the partitioning of data. For this, 80% of the dataset is considered for training and 20% for testing. Also, 20% of training data is used as the validation set. The validation set is used to avoid overfitting of the DL models. Each of the experiments was performed 10 times, and the average and standard deviation are reported here. The settings of the parameters of the DL methods are given in Table 4. The parameters' values are determined empirically.

Table 4: Settings of DL algorithms

| Parameter | Value |
|---|---|
| Activation function | Sigmoid |
| Loss | Binary_crossentropy |
| Optimizer | adam |
| Batch_size | 64 |
| Epochs | 3 |
| Validation_split | 0.2 |

## B) Performance Analysis

The results of DL and voting methods on the PURE dataset are given in Table 5 where the Keras word embedding method is employed. The performance of each method is given in terms of Precision, Recall, F1-Score, and standard deviation. All the DL methods show competitive performance. However, for three metrics precision, recall, and F1-score, the BiLSTM model obtains the first rank among DL models. The last rank belong to GRU models. It is apparent from the table that the voting methods improve the performance and the best results are obtained by the hard-voting method.

Table 5: The results of deep learning methods on the PURE dataset using the Keras word embedding method

| | Precision (Std. Dev.) | Recall (Std. Dev.) | F-score (Std. Dev.) |
|---|---|---|---|
| LSTM | 79.06(1.29) | 79.09(1.29) | 79.04(1.31) |
| BiLSTM | 79.49(1.34) | 79.38(1.52) | 79.32(1.47) |
| GRU | 78.61(1.45) | 78.37(1.44) | 78.31(1.42) |
| BiGRU | 79.27(1.07) | 79.25(1.14) | 79.23(1.13) |
| CNN | 78.83(2.30) | 78.39(2.54) | 78.21(2.68) |
| Hard voting | 80.16(1.32) | 80.12(1.39) | 80.08(1.36) |
| Soft voting | 80.14(1.15) | 80.11(1.20) | 80.06(1.18) |

The results of DL and voting methods on the PURE dataset are given in Table 6 using the GloVe word embedding method. The DL methods show competitive results. Among the DL models, the first rank is obtained by the CNN models while the BiGRU placed at the last rank. The ranking is identical over three metrics precision, recall, and F-score. Similar to the previous experiment, the voting methods improve the performance of DL models. The best results were obtained by the Hard voting method. The results show that DL methods produce better results when they use Keras word embedding.

Table 6: The results of deep learning methods on the PURE dataset using the GloVe embedding method

| | Precision (Std. Dev.) | Recall (Std. Dev.) | F-score (Std. Dev.) |
|---|---|---|---|
| LSTM | 73.28(1.30) | 72.82(1.39) | 72.80(1.39) |
| BiLSTM | 73.00(1.39) | 72.55(1.18) | 72.54(1.24) |
| GRU | 73.16(1.03) | 72.91(0.98) | 72.88(1.03) |
| BiGRU | 72.85(1.14) | 72.50(0.99) | 72.47(1.09) |
| CNN | 73.55(1.24) | 73.28(1.48) | 73.06(1.57) |
| Hard voting | 74.17(1.22) | 73.97(1.21) | 73.92(1.25) |
| Soft voting | 73.76(0.92) | 73.36(0.87) | 73.49(0.96) |

In summary, hard and soft voting algorithms have succeeded to improve the results. Among the two voting algorithms, hard voting has performed slightly better than soft voting. When the default word embedding method of the Keras library was used for feature extraction, the results are better. These better results come from the fact that in the first case, the word embedding vector is trained on the dataset at hand, and it recognizes the similarity and connection between the words of the dataset more accurately.

## 5. Conclusions

Requirement engineering is the most important phase of software production, and the use of deep learning methods in most engineering fields automates the process and increases accuracy. In this work, a method based on DL models was presented to classify software requirements automatically. The performance of DL methods for requirement classification was analyzed on the PURE repository. The results showed that the suggested methods were efficient and they can successfully classify the requirements. Five DL algorithms and two voting classification algorithms were used, and this proposed method was tested on PURE datasets. Experiments showed that the use of voting mechanisms has increased the accuracy of DL methods. On the other hand, the proposed method was also tested with Keras and GloVe embedding methods, and the results show that the default word embedding method of the Keras library works better than the GloVe pre-trained word embedding method.

## References

[1] R. Sepahvand, R. Akbari, S. Hashemi, and O. Boushehrian, 2022. An Effective Model to Predict the Extension of Code Changes in Bug Fixing Process Using Text Classifiers. Iranian Journal of Science and Technology, Transactions of Electrical Engineering, 46(1), pp.257-274.

[2] Z. Javidi, R. Akbari,O. Bushehrian, 2021. A new method based on formal concept analysis and metaheuristics to solve class responsibility assignment problem. Iran Journal of Computer Science, 4(4), pp.221-240.

[3] V. Etemadi, O. Bushehrian, R. Akbari, and G. Robles, 2021. A scheduling-driven approach to efficiently assign bug fixing tasks to developers. Journal of Systems and Software, 178, p.110967.

[4] M. Yousefi, R. Akbari, and S. M. R. Moosavi. "Using Machine Learning Methods for Automatic Bug Assignment to Developers." Journal of Electrical and Computer Engineering Innovations (JECEI) 8, no. 2 (2020): 263-272.

[5] R. Sepahvand, R. Akbari, and S. Hashemi, 2020. Predicting the bug fixing time using word embedding and deep long short term memories. IET Software, 14(3), pp.203-212.

[6] R. N. Almanza, R. J. Ramirez , G. Licea, "Towards supporting Software Engineering using Deep Learning: A case of Software Requirements Classification," in 5th International Conference in Software Engineering Research and Innovation (CONISOFT), 2017.

[7] M. A. Rahman, M. A. Haque, M. N. A. Tawhid , M. S. Siddik, "Classifying Non-functional Requirements using RNN Variants," in Proceedings of the 3rd ACM SIGSOFT International Workshop on Machine Learning Techniques for Software Quality Evaluation, 2019.

[8] J. Winkler , A. Vogelsang, "Automatic Classification of Requirements Based on Convolutional Neural Networks," in IEEE 24th International Requirements Engineering Conference Workshops, 2016.

[9] C. Baker, L. Deny, S. Chakraborty , J. Dehlinger, "Automatic multi-class non-functional software requirements classification using neural networks," in IEEE 43rd Annual Computer Software and Applications Conference (COMPSAC), 2019.

[10] N. Rahimi, F. Eassa, L. Elrefaei, "One-and two-phase software requirement classification using ensemble deep learning". Entropy, 23(10), 1264. 2021.

[11] D.Kici, G.Malik, M.Cevik, D.Parikh, A. Basar, "A BERT-based transfer learning approach to text classification on software requirements specifications". In Canadian Conference on AI.2021.

[12] V. Ivanov, A. Sadovykh, A. Naumchev, A. Bagnato, K.Yakovlev,. "Extracting Software Requirements from Unstructured Documents". arXiv preprint arXiv:2202.02135. 2022.

[13] G. Malik., M.Cevik, Y. Khedr, D. Parikh, A. Basar. "Named Entity Recognition on Software Requirements Specification Documents". In Canadian Conference on AI. 2021.

[14] N. Rahimi, F. Eassa, L. Elrefaei. "An ensemble machine learning technique for functional requirement classification." symmetry, 12(10), 1601. 2020

[15] S. Tiun, U. A. Mokhtar, S. H. Bakar, S. Saad. "Classification of functional and non-functional requirement in software requirement using Word2vec and fast Text". In journal of Physics: conference series (Vol. 1529, No. 4, p. 042077). IOP Publishing. 2020, April.

[16] M. A. Haque, M. A. Rahman and M. S. Siddik, "Non-Functional Requirements Classification with," in 1st International Conference on Advances in Science, Engineering and Robotics Technology (ICASERT), 2019.

[17] Z. Shakeri Hossein Abad, O. Karras, P. Ghazi, M. Glinz, G. Ruhe and K. Schneider, "What Works Better? A Study of Classifying Requirements," in IEEE 25th International Requirements Engineering Conference, 2017.

[18] M. Lu, P. Liang, "Automatic Classification of Non-Functional Requirements from," in 21st International Conference on Evaluation and Assessment in Software Engineering, 2017.

[19] R. Jindal, R. Malhotra., A. Jain, "Automated classification of security requirements." In 2016 International Conference on Advances in Computing, Communications and Informatics (ICACCI), pp. 2027-2033. IEEE, 2016.

[20] E. Dias Canedo, Edna, B. Cordeiro Mendes. "Software Requirements Classification Using Machine Learning Algorithms." Entropy 22, no. 9 (2020): 1057.

[21] R. Deocadez, R. Harrison, D. Rodriguez. "Automatically classifying requirements from app stores: A preliminary study." In 2017 IEEE 25th International Requirements Engineering Conference Workshops (REW), pp. 367-371. IEEE, 2017.

[22] P. Talele, R. Phalnikar. "Classification and Prioritisation of Software Requirements using Machine Learning–A Systematic Review." In 2021 11th International Conference on Cloud Computing, Data Science & Engineering (Confluence), pp. 912-918.

[23] L. Tóth, L. Vidács. "Study of various classifiers for identification and classification of non-functional requirements." In International Conference on

Computational Science and Its Applications, pp. 492-503. Springer, Cham, 2018.

[24] L. Tóth, L. Vidács. "Comparative Study of The Performance of Various Classifiers in Labeling Non-

Functional Requirements." Information Technology and Control 48, no. 3 (2019): 432-445.

[25] M. Anas, G. Williams. "Detecting, classifying, and tracing non-functional software requirements." Requirements Engineering 21, no. 3 (2016): 357-381.